\documentclass{elsart}
\usepackage{amssymb}

\begin{document}

\begin{frontmatter}

\title{An alternate mathematical model for single-wall carbon nanotubes}

\author{Nicolae Cotfas}

\address{Faculty of Physics, University of Bucharest,
PO Box 76-54, Postal Office 76, Bucharest, Romania}
\ead{ncotfas@yahoo.com}
\begin{abstract}
The positions of atoms forming a carbon nanotube are usually described by using 
a system of generators of the symmetry group. Each atomic position 
corresponds to an element of the set 
$\mathbb{Z}\times \{ 0,1,...,n\} \times \{ 0,1\}$, where $n$ is a natural
number depending on the considered nanotube.
We obtain an alternate rather different description by starting from 
a description of the honeycomb lattice in terms of Miller indices.
In our mathematical model which is a factor space defined by an 
equivalence relation in the set 
$\left\{ (v_0,v_1,v_2)\in \mathbb{Z}^3\ |\ v_0+v_1+v_2\in \{ 0,1\}\right\}$
the neighbours of an atomic position can be described in a simpler way,
and the mathematical objects with geometric or physical significance have
a simpler and more symmetric form.
\end{abstract}


\end{frontmatter}

\section{Introduction}

The carbon nanotubes, discovered by Iijima in 1991, 
have several remarkable physical properties (geometry-dependent electronic
transport from metallic to semiconducting with narrow and moderate band gaps,
record-high elastic modulus, light weight) and many potential
applications (molecular electronic devices, fiber reinforcement technologies,
flat displays, carbon-based nanotips). 
Extensive experimental and theoretical investigations have been carried out 
on the mechanical and electronic properties of these novel fibers. 

The structure of a single-wall carbon nanotube observed by scanning tunneling 
microscopy \cite{V} can be visualized as the structure obtained by rolling a
graphene sheet such that the endpoints of a translation vector
are folded one onto the other. The geometric and physical properties
of the obtained carbon nanotube depend on this vector, called the {\it chirality} 
of the tubule. The position of the atoms forming a carbon nanotube are usually
described by using a system of generators of the corresponding symmetry group.
Our purpose is to present an alternate mathematical model obtained by starting
from a three-axes description (that is, a description in terms of Miller indices 
\cite{S}) of the honeycomb lattice.

\section{Honeycomb lattice in a three-axes description}

The vectors $e_0=\left( 2/\sqrt{6},0\right)$,
$e_1=\left( -1/\sqrt{6},1/\sqrt{2}\right)$,
$e_2=\left( -1/\sqrt{6},-1/\sqrt{2}\right)$
corresponding to the vertices of an equilateral triangle form a system
of coherent vectors \cite{C0}, that is, for any  
$v=(v_x,v_y),\ u=(u_x,u_y)\in \mathbb{R}^2$ we have
\begin{equation} 
v=\sum_{i=0}^2\langle v,e_i\rangle e_i\qquad 
\langle v,u\rangle =\sum_{i=0}^2\langle v,e_i\rangle \langle u,e_i\rangle \qquad 
||v||^2=\sum_{i=0}^2\langle v,e_i\rangle ^2 
\end{equation}
where $\langle ,\rangle $ is the usual scalar product.
The numbers $\tilde{v}_0=\langle v,e_0\rangle $, 
$\tilde{v}_1=\langle v,e_1\rangle $, 
$\tilde{v}_2=\langle v,e_2\rangle $ satisfy the 
relation $\tilde{v}_0+\tilde{v}_1+\tilde{v}_2=0$ and
can be regarded as the {\it canonical coordinates} of $v$ with respect to 
the system of vectors $\{ e_0,e_1,e_2\}$. The space
\begin{equation}
{\mathcal E}=\{ (u_0,u_1,u_2)\in \mathbb{R}^3\ |\ u_0+u_1+u_2=0\}
\end{equation}
becomes in this way a mathematical model for the geometric plane.
The correspondence between this description and the usual one is given 
by the isometry 
\begin{equation}
\mathcal{I}:\mathbb{R}^2\longrightarrow \mathcal{E}\qquad 
\mathcal{I} v=(\tilde{v}_0,\tilde{v}_1,\tilde{v}_2)\qquad 
\mathcal{I}^{-1}(u_0,u_1,u_2)=\sum_{i=0}^2u_ie_i.
\end{equation}
 
The representation of a vector $v\in \mathbb{R}^2$ as a linear 
combination of $e_0,\ e_1,\ e_2$ is not unique. More exactly, we have
\[
v=\sum_{i=0}^2v_ie_i\qquad \Longleftrightarrow \qquad 
(v_0,v_1,v_2)\in \{ (\tilde{v}_0+\alpha ,\tilde{v}_1+\alpha ,\tilde{v}_2+\alpha )\ |\
\alpha \in \mathbb{R}\}.
\]
For each $v\in \mathbb{R}^2$ we denote by $(v_0,v_1,v_2)$ (or simply by $v$)
an  element of $\mathbb{R}^3$ such that $v=v_0e_0+v_1e_1+v_2e_2$. 
One can verify by direct computation that
\begin{equation}
\langle u,v\rangle =\sum_{i=0}^2\tilde{u}_i\tilde{v}_i
=\sum_{i=0}^2\tilde{u}_iv_i
=\sum_{i=0}^2u_i\tilde{v}_i
\end{equation}
for any $u,v\in \mathbb{R}^2$.

\begin{figure}
\setlength{\unitlength}{1mm}
\begin{picture}(110,50)(-10,0)
\multiput(0,0)(18,0){7}{\line(1,0){6}}
\multiput(18,0)(18,0){6}{\line(-2,3){3}}
\multiput(9,4.5)(18,0){6}{\line(-2,-3){3}}
\multiput(9,4.5)(18,0){6}{\line(1,0){6}}
\multiput(9,4.5)(18,0){6}{\line(-2,3){3}}
\multiput(18,9)(18,0){6}{\line(-2,-3){3}}
\multiput(0,9)(18,0){7}{\line(1,0){6}}
\multiput(18,9)(18,0){6}{\line(-2,3){3}}
\multiput(9,13.5)(18,0){6}{\line(-2,-3){3}}
\multiput(9,13.5)(18,0){6}{\line(1,0){6}}
\multiput(9,13.5)(18,0){6}{\line(-2,3){3}}
\multiput(18,18)(18,0){6}{\line(-2,-3){3}}
\multiput(0,18)(18,0){7}{\line(1,0){6}}
\multiput(18,18)(18,0){6}{\line(-2,3){3}}
\multiput(9,22.5)(18,0){6}{\line(-2,-3){3}}
\multiput(9,22.5)(18,0){6}{\line(1,0){6}}
\multiput(9,22.5)(18,0){6}{\line(-2,3){3}}
\multiput(18,27)(18,0){6}{\line(-2,-3){3}}
\multiput(0,27)(18,0){7}{\line(1,0){6}}
\multiput(18,27)(18,0){6}{\line(-2,3){3}}
\multiput(9,31.5)(18,0){6}{\line(-2,-3){3}}
\multiput(9,31.5)(18,0){6}{\line(1,0){6}}
\multiput(9,31.5)(18,0){6}{\line(-2,3){3}}
\multiput(18,36)(18,0){6}{\line(-2,-3){3}}
\multiput(0,36)(18,0){7}{\line(1,0){6}}
\multiput(18,36)(18,0){6}{\line(-2,3){3}}
\multiput(9,40.5)(18,0){6}{\line(-2,-3){3}}
\multiput(9,40.5)(18,0){6}{\line(1,0){6}}
\multiput(9,40.5)(18,0){6}{\line(-2,3){3}}
\multiput(18,45)(18,0){6}{\line(-2,-3){3}}
\multiput(0,45)(18,0){7}{\line(1,0){6}}
\multiput(18,45)(18,0){6}{\line(-2,3){3}}
\multiput(9,49.5)(18,0){6}{\line(-2,-3){3}}
\multiput(9,49.5)(18,0){6}{\line(1,0){6}}
\put(18,9){\vector(4,1){36}}
\thinlines
\put(18,9){\line(-1,4){10}}
\put(54,18){\line(-1,4){8}}
\put(18,9){\line(1,-4){2.5}}
\put(54,18){\line(1,-4){4.5}}
\put(90,27){\line(-1,4){6}}
\put(90,27){\line(1,-4){7}}
\put(54,18){\vector(2,1){9}}
\put(54,18){\vector(2,-1){9}}
\put(54,19){$O$}
\put(90,27.5){$A$}
\put(58,22){$a_2$}
\put(58,13){$a_1$}
\put(51.6,12.5){$e_2$}
\put(48.1,20){$e_1$}
\put(60.5,17){$e_0$}
\put(84,26){$c$}
\put(13.5,8){$-\!c$}
\put(61.5,35){$v$}
\put(25,26){$v\!\!-\!\!c$}
\put(97,44){$v\!\!+\!\!c$}
\put(54,36.5){$v^{\!1}$}
\put(63,41){$v^{\!3}$}
\put(63,28.5){$v^{\!2}$}
\put(54,18){\circle*{1}}
\put(90,27){\circle*{1}}
\thicklines
\put(54,18){\vector(4,1){36}}
\put(54,18){\vector(1,0){6}}
\put(54,18){\vector(-2,3){3}}
\put(54,18){\vector(-2,-3){3}}
\end{picture}
\caption{The honeycomb lattice and the partition defined by a vector
$c\in {\mathcal T}$.}
\end{figure}
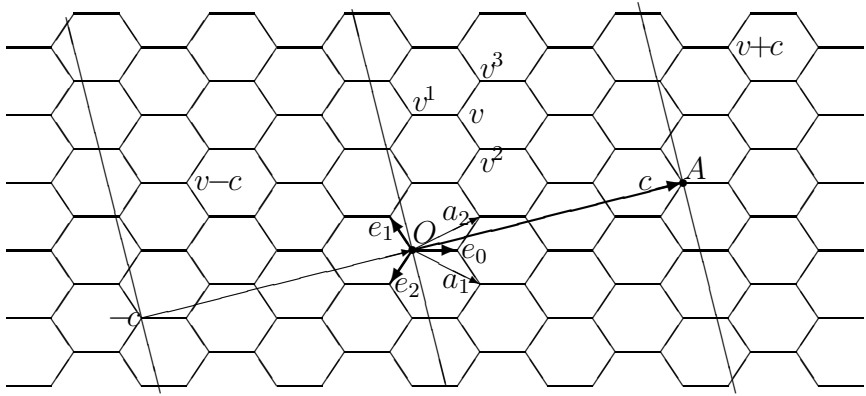

The points of the plane $\mathbb{R}^2$ corresponding to the elements of the set
\begin{equation}
{\mathcal L}=\left\{ v=(v_0,v_1,v_2)\in \mathbb{Z}^3\ |\ v_0+v_1+v_2\in \{ 0,1\}\right\} 
\end{equation}
are distinct and form \cite{C1} the {\it honeycomb lattice}
\begin{equation}
\Lambda =\left\{ \left. \sum_{i=0}^2v_ie_i\ \right|\ 
v=(v_0,v_1,v_2)\in {\mathcal L}\right\}.
\end{equation}
The bijection ${\mathcal L}\longrightarrow \Lambda :v\mapsto \sum_{i=0}^2v_ie_i$
allows us to describe $\Lambda $ by using ${\mathcal L}$.

The {\it nearest neighbours} of $v$ are
\[ v^0=(v_0+\nu (v),v_1,v_2)\quad v^1=(v_0,v_1+\nu (v),v_2)\quad
v^2=(v_0,v_1,v_2+\nu (v))\]
where $\nu (v)=(-1)^{v_0+v_1+v_2}$. The six points $v^{ij}=(v^i)^j$
corresponding to $i\not=j$ are the {\it next-to-nearest neighbours} of $v$, 
and one can remark that $v^{ii}=v$, $v^{ijl}=v^{lji}$, 
for any $i,\, j,\, l\in \{ 0,1,2 \}$. The mapping \cite{C2}
\begin{equation} 
d:{\mathcal L}\times {\mathcal L}\longrightarrow \mathbb{N}\qquad 
d(v,u)=|v_0-u_0|+|v_1-u_1|+|v_2-u_2|
\end{equation}
is a distance on ${\mathcal L}$, and a point $u$ is a {\it neighbour of order} $l$
of $v$ if $d(v,u)=l$.

We have ${\mathcal L}={\mathcal T}\cup ({\mathcal T}+\vartheta )$, 
where $\vartheta =(1,0,0)$ and 
\begin{equation}
{\mathcal T}=\{ v=(v_0,v_1,v_2)\in \mathbb{Z}^3\ |\ v_0+v_1+v_2=0\} =
{\mathcal E}\cap \mathbb{Z}^3.
\end{equation} 
The symmetry group $G$ of the honeycomb lattice coincides with the group of
all the isometries of the metric space $({\mathcal L},d)$, and is generated by
the transformations \cite{C2}
\begin{equation}\begin{array}{l}
\sigma :{\mathcal L}\longrightarrow {\mathcal L}\qquad \sigma (v_0,v_1,v_2)=(v_1,v_2,v_0)\\
\varrho :{\mathcal L}\longrightarrow {\mathcal L}\qquad \varrho (v_0,v_1,v_2)=(v_0,v_2,v_1)\\
\tau :{\mathcal L}\longrightarrow {\mathcal L}\qquad \tau (v_0,v_1,v_2)
=-(v_0,v_1,v_2)+\vartheta .
\end{array}
\end{equation}
The group $G$ contains the subgroup of translations (also denoted by $\mathcal {T}$)
\[ \{{\mathcal L}\longrightarrow {\mathcal L}:v\mapsto v+u \ |\ u\in \mathcal {T}\ \} \]
generated by $\sigma ^2\tau \sigma \tau $ and $\sigma \tau \sigma ^2\tau $
\begin{equation}\begin{array}{l}
\sigma ^2\tau \sigma \tau (v_0,v_1,v_2)=(v_0,v_1,v_2)+(-1,1,0)\\
\sigma \tau \sigma ^2\tau (v_0,v_1,v_2)=(v_0,v_1,v_2)+(-1,0,1).
\end{array}
\end{equation}

It is known \cite{S2} that the C-C bond length in the case of a graphene sheet is 1.44 \AA.
If we use the honeycomb lattice ${\mathcal L}$ as a mathematical model for a 
graphene sheet then we have to multiply the position vectors of the points
of the lattice by the constant $a=1.44\sqrt{6}/2$  in order to get their lengths in \AA.

Consider the Hilbert space
$(l^2({\mathcal L}),\langle , \rangle )$, where
\[
l^2({\mathcal L})=\left\{ \psi :{\mathcal L}\longrightarrow \mathbb{C}\ \left| \ 
\sum_{v\in {\mathcal L}}|\psi (v)|^2<\infty \right. \right\} \]
\begin{equation}
\langle \psi _1,\psi _2\rangle =
\sum_{v\in {\mathcal L}}\overline{\psi }_1(v)\psi _2(v) 
\end{equation}
and the unitary representation of $G$ in $l^2({\mathcal L})$ defined by 
\begin{equation}
g:l^2({\mathcal L})\longrightarrow l^2({\mathcal L})\qquad 
(g\psi )(v)=\psi (g^{-1}v).
\end{equation}

Let $\varepsilon $ be a real number, and  $\gamma _0, \gamma _1,\gamma _2$ be 
three complex numbers. The linear operator
\begin{equation}\label{H}
H:l^2({\mathcal L})\longrightarrow l^2({\mathcal L})\qquad
(H\psi )(v)=\varepsilon \psi (v)+\sum_{j=0}^2\gamma (v,v^j)\, \psi (v^j)
\end{equation}
where
\begin{equation}\label{gamma}
\gamma (v,v^j)=\left\{ \begin{array}{lll}
\gamma _j & {\rm if} & \nu (v)=1\\
\overline{\gamma }_j & {\rm if} & \nu (v)=-1
\end{array} \right. 
\end{equation}
is a self-adjoint operator
\[ \langle H\psi _1,\psi _2\rangle =
\varepsilon \sum_{v\in {\mathcal L}}
\overline{\psi }_1(v)\psi _2(v)+\sum_{j=0}^2\sum_{v\in {\mathcal L}}
\overline{\gamma }(v,v^j)\overline{\psi }_1(v^j)\psi _2(v)\]
\[=\varepsilon \sum_{v\in {\mathcal L}}
\overline{\psi }_1(v)\psi _2(v)+\sum_{j=0}^2\sum_{v\in {\mathcal L}}
{\gamma }(v,v^j)\overline{\psi }_1(v)\psi _2(v^j)
= \langle \psi _1,H\psi _2\rangle .\]
The Hamiltonian used in the tight-binding description of $\pi $ bands
in 2D graphite, with only first-neighbour $C-C$ interaction, has the
form (\ref{H}).\\[2mm]
{\bf Theorem 1.} {\it For any $k=(k_0,k_1,k_2)\in \mathcal{E}$ the real numbers
\begin{eqnarray}\label{roots}
E_{\pm }(k)=\varepsilon \pm
|\gamma _0 {\rm e}^{ {\rm i}k_0a}+\gamma _1  {\rm e}^{ {\rm i}k_1a}+\gamma _2  
{\rm e}^{ {\rm i}k_2a}| 
\end{eqnarray}
belong to the spectrum of} $H.$\\[2mm]
{\bf Proof.} The function
\[ \varphi :\mathcal{L}\longrightarrow \mathbb{C}\qquad
\varphi (v)=\left\{ \begin{array}{lll}
\alpha \ \ & {\rm for\ \ } & v\in \mathcal{T}\\
\beta & {\rm for} & v\in \mathcal{T}+\vartheta 
\end{array}\right. \]
where $\alpha ,\, \beta $ are two constants, is invariant under any translation
$u\in \mathcal{T}$ 
\[ \varphi (v+u)=\varphi (v)\qquad {\rm for\ all} \quad v\in \mathcal{L}.\]
The Bloch type function
\begin{equation}
\psi _k:\mathcal{L}\longrightarrow \mathbb{C}\qquad 
\psi _k(v)=\varphi (v)\, {\rm e}^{{\rm i}\langle k,v\rangle a}
\end{equation}
belonging to an extension of the space $l^2(\mathcal{L})$ satisfies the relation 
$H\psi _k=E\psi _k$ if and only if $(\alpha ,\beta )$ is a solution of the 
system of equations
\[ \left\{ \begin{array}{l}
\varepsilon \alpha +(\gamma _0 {\rm e}^{ {\rm i}k_0a}+\gamma _1  {\rm e}^{ {\rm i}k_1a}+
\gamma _2  {\rm e}^{ {\rm i}k_2a})\beta =E\alpha \\
(\overline{\gamma }_0 {\rm e}^{- {\rm i}k_0a}+\overline{\gamma }_1 
 {\rm e}^{- {\rm i}k_1a}+\overline{\gamma }_2  {\rm e}^{- {\rm i}k_2a})\alpha +\varepsilon \beta=E\beta .
\end{array} \right. \]
This system has non-trivial solutions if and only if
\[ \left| \begin{array}{cc}
\varepsilon -E & \gamma _0 {\rm e}^{ {\rm i}k_0a}+\gamma _1  {\rm e}^{ {\rm i}k_1a}+
\gamma _2  {\rm e}^{ {\rm i}k_2a}\\
\overline{\gamma }_0 {\rm e}^{- {\rm i}k_0a}+\overline{\gamma }_1  {\rm e}^{- {\rm i}k_1a}+
\overline{\gamma }_2  {\rm e}^{- {\rm i}k_2a} & \varepsilon -E
\end{array} \right| =0\]
that is, if and only if $E$ is one of the numbers
$E_{\pm }(k) $.\qquad  \rule{2mm}{2mm}

The origin on the energy axis is usually chosen such that $\varepsilon =0$.
If $\gamma _0=\gamma _1=\gamma _2=\gamma $ is a real positive number then
$H$ is a $G$-invariant self-adjoint operator and its spectrum contains 
for each $k\in \mathcal {E}$ the numbers $\pm E(k)$, where
\[ E(k)=\gamma | {\rm e}^{ {\rm i}k_0a}+ {\rm e}^{ {\rm i}k_1a}+ {\rm e}^{ {\rm i}k_2a}|\]
\begin{equation} \label{Ek}
\mbox{}\quad \ \ \ =\gamma  \sqrt{3+2\cos(k_0-k_1)a+
2\cos(k_1-k_2)a+2\cos(k_2-k_0)a}.
\end{equation} 

\begin{figure}
\setlength{\unitlength}{1mm}
\begin{picture}(60,40)(-10,2)
\put(30,0){\line(3,2){18}}
\put(30,0){\line(-3,2){18}}
\put(48,12){\line(0,1){20}}
\put(12,12){\line(0,1){20}}
\put(12,32){\line(3,2){18}}
\put(48,32){\line(-3,2){18}}
\put(30,22){\vector(1,0){25}}
\put(30,22){\vector(-2,3){14}}
\put(30,22){\vector(-2,-3){14}}
\put(30,22){\circle*{0.8}}
\put(48,22){\circle*{0.8}}
\put(48,12){\circle*{0.8}}
\put(30,23){$\Gamma $}
\put(43,19){$M$}
\put(48.5,10){$K$}
\put(114,28){$c$}
\put(92,31){$\omega $}
\put(75,45){$b$}
\put(80,22){\vector(4,1){36}}
\put(80,22){\vector(2,1){15}}
\put(80,22){\vector(-1,4){6}}
\put(80.1,22){\line(-1,4){5}}
\put(80.2,22){\line(-1,4){5}}
\put(85,15){\line(-1,4){5}}
\put(85.1,15){\line(-1,4){5}}
\put(85.2,15){\line(-1,4){5}}
\put(90,8){\line(-1,4){5}}
\put(90.1,8){\line(-1,4){5}}
\put(90.2,8){\line(-1,4){5}}
\put(95,1){\line(-1,4){5}}
\put(95.1,1){\line(-1,4){5}}
\put(95.2,1){\line(-1,4){5}}
\end{picture}
\caption{The first Brillouin zone ${\mathcal B}$ (left) and the set
${\mathcal B}_c$ in case $n=4$ (right).}
\end{figure}
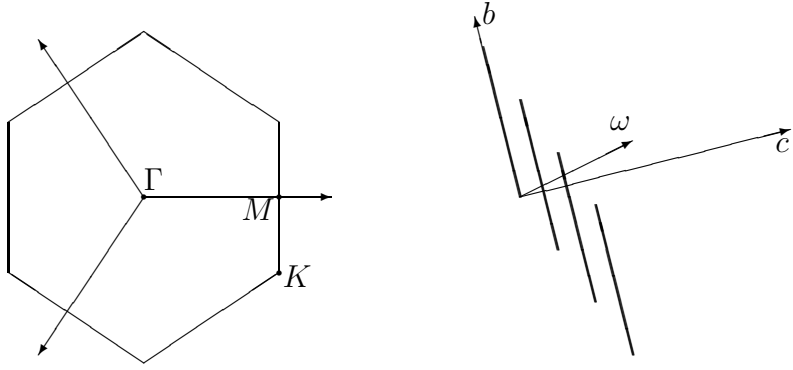

The relation (\ref{Ek}) allows us to extend the function 
$E:\mathcal{E}\longrightarrow \mathbb{R}$ to a periodic function
$E:\mathbb{R}^3\longrightarrow \mathbb{R}$ 
\begin{eqnarray}
E(k_0,k_1,k_2) &=& E\left(k_0+\frac{2\pi }{a},k_1,k_2\right)\nonumber \\
&=& E\left(k_0,k_1+\frac{2\pi }{a},k_2\right)
=E\left(k_0,k_1,k_2+\frac{2\pi }{a}\right) 
\end{eqnarray}
with the property
\[ E(k_0,k_1,k_2)=E(k_0+\alpha ,k_1+\alpha ,k_2+\alpha )\qquad  
{\rm for\ all\ } \alpha \in \mathbb{R}.\]
The corresponding first Brillouin zone is the hexagonal set (figure 2) 
\begin{equation}
 {\mathcal B}=
\left\{ (k_0,k_1,k_2)\in \mathcal{E}\ \left| \ 
-\frac{2\pi }{3a}\!\leq \!k_i\!\leq \!\frac{2\pi }{3a} \right. \right\}
\end{equation}
(certain points lying on the frontier of ${\mathcal B}$ are equivalent).

The intervals $[-3\gamma ,0]=\{ -E(k)\ |\ k\in {\mathcal B}\}$ and 
$[0,3\gamma ]=\{ E(k)\ |\ k\in {\mathcal B}\}$ correspond to the valence 
$\pi $ and the conduction $\pi ^*$ energy bands, respectively.
The graphene sheet is a conductor since the gap between these bands is null.

Since $E(k)$ can be written as 
\[ E(k)\!=\!\gamma \sqrt{(\cos \, k_0a\!+\!\cos \, k_1a\!+\!\cos \, k_2a)^2\!+\!
(\sin \, k_0a\!+\!\sin \, k_1a\!+\!\sin \, k_2a)^2}\]
we have $E(k)\geq 0$, and the only points of $\mathcal {B}$ with $E(k)=0$ are
\begin{equation} 
\pm \left( \frac{2\pi }{3a},-\frac{2\pi }{3a},0\right),
\pm \left( \frac{2\pi }{3a},0,-\frac{2\pi }{3a}\right),
\pm \left( 0,\frac{2\pi }{3a},-\frac{2\pi }{3a}\right)
\end{equation} 
that is, the vertices of the Brillouin zone (usually denoted by $K$) \cite{Mi}.

It is known that the Fermi level for a graphene sheet occurs at the $K$
points. The function $E:\mathbb{R}^3\longrightarrow \mathbb{R}$ is not 
differentiable at these points. We have, for example,
\begin{eqnarray}  \lim_{k_0\rightarrow \frac{2\pi }{3a}}
\frac{E\left(k_0,-\frac{2\pi }{3a},0\right)-0}{k_0-\frac{2\pi }{3a}} &=&
\gamma \lim_{k_0\rightarrow \frac{2\pi }{3a}}
\frac{\sqrt{2-2\cos\left(k_0-\frac{2\pi }{3a}\right)}}{k_0-\frac{2\pi }{3a}}\nonumber \\
&=& 2\gamma \lim_{k_0\rightarrow \frac{2\pi }{3a}}
\frac{\left| \sin \frac{k_0-(2\pi /3a)}{2}\right|}{k_0-\frac{2\pi }{3a}} 
\end{eqnarray}
whence
\[  \lim_{{\scriptsize \begin{array}{l} k_0\rightarrow 2\pi /3a\\
                                      k_0>2\pi /3a\end{array}}}
\frac{E\left(k_0,-\frac{2\pi }{3a},0\right)-0}{k_0-\frac{2\pi }{3a}} =\gamma 
 \qquad \lim_{{\scriptsize \begin{array}{l} k_0\rightarrow 2\pi /3a\\
                                      k_0<2\pi /3a\end{array}}}
\frac{E\left(k_0,-\frac{2\pi }{3a},0\right)-0}{k_0-\frac{2\pi }{3a}} =
-\gamma .\]

The function $E:\mathbb{R}^3\longrightarrow \mathbb{R}$ is differentiable at any point 
$k$ with $E(k)\not=0$, and 
\begin{equation} 
 \frac{\partial E}{\partial k_0}(k)=\gamma \frac
{-a\sin (k_0-k_1)a+a\sin (k_2-k_0)a}{
\sqrt{3\!+\!2\cos(k_0\!-\!k_1)a\!+\!2\cos(k_1\!-\!k_2)a\!+\!2\cos(k_2\!-\!k_0)a}} 
\end{equation} 
etc. The stationary points lying in the Brillouin zone $\mathcal{B}$ are
\[  
(0,0,0),\ \pm \left(\frac{2\pi }{3a},-\frac{\pi }{3a},-\frac{\pi }{3a}\right),\ 
\pm \left(-\frac{\pi }{3a},\frac{2\pi }{3a},-\frac{\pi }{3a}\right),\ 
\pm \left(-\frac{\pi }{3a},-\frac{\pi }{3a},\frac{2\pi }{3a}\right)
\] 
that is, the center (a maximum point, denoted by $\Gamma $) and the 
middle of the edges of the Brillouin zone (saddle points, denoted by $M$).

\section{Chiral single-wall carbon nanotubes}

A single-wall carbon nanotube can be visualised as the structure obtained by 
rolling a graphene sheet such that the endpoints $O$ and $A$ of a translation
vector $\vec{OA}$ are folded one onto the other (figure 1).
The geometric and physical properties of the obtained nanotube depend on this
vector, called the {\em chirality} of the tubule and represented in our approach 
by an element $c\in \mathcal {T}$.
Without loss of generality, we can restrict us to the tubules with
$c_0>c_1\geq c_2.$ In the case $c_1=c_2$ we have an {\it armchair} tubule, 
and in the case $c_1=0$ a {\it zig-zag} tubule. The nanotubes with 
$0\not=c_1\not=c_2$ are called {\em chiral nanotubes}. The diameter of the nanotube
of chirality $c$ is $||c||a/\pi $.

After the graphene sheet 
rolling, the points $ ...,\ v-2c,\ v-c,\ v,\ v+c,\ v+2c,\ ... $
are folded one onto the other, for any $v=(v_0,v_1,v_2)\in {\mathcal L}.$ 
Thus, each point of the set
\begin{equation} 
[v_0,v_1,v_2]=v+\mathbb{Z}c=
\{ \ (v_0+jc_0,v_1+jc_1,v_2+jc_2)\ |\ \ j\in \mathbb{Z}\ \} 
\end{equation} 
describes the same point of the carbon nanotube of chirality $c.$
Each rational number is a class of equivalent fractions, 
called its representatives. In a similar way, for each point of a carbon
nanotube we have an infinite number of possibilities to describe it in
our model. A mathematical expression containing the coordinates of a
point is well-defined only if it does not depend on the representative 
we choose. We describe the atomic positions on a carbon nanotube by using
the subset
\begin{equation}
{\mathcal L}_c=\left\{ \ [v_0,v_1,v_2]\in \mathbb{Z}^3/(\mathbb{Z}c)\ \ | 
\ \ \ v_0+v_1+v_2\in \{ 0;1\} \ \ \right\} . 
\end{equation}
of the factor space
\begin{equation}
 \mathbb{Z}^3/(\mathbb{Z}c)=
\{ \ [v]=(v_0,v_1,v_2)+\mathbb{Z}c\ |\  v_0,\ v_1,\ v_2\in \mathbb{Z}\ \}
\end{equation}
as a mathematical model. 
One can remark that $\mathcal{L}_c$ is well defined since the condition 
$v_0+v_1+v_2\in \{ 0;1\}$ we impose to $[v_0,v_1,v_2]$ does not depend on
the representative we choose. Indeed, 
$(v_0+jc_0)+(v_1+jc_1)+(v_2+jc_2)=v_0+v_1+v_2$, for all $j\in \mathbb{Z}.$ 

Each point $[v]\in {\mathcal L}_c$ has three {\it nearest 
neighbours}, namely, $[v^0],$ $[v^1],$ $[v^2],$
and six {\it next-to-nearest neighbours}, namely, 
$[v^{01}],$ $[v^{10}],$ $[v^{12}],$ $[v^{21}],$ $[v^{20}],$
$[v^{02}].$

A symmetry transformation of the honeycomb lattice
${\mathcal L}\longrightarrow {\mathcal L}:v\mapsto gv$ 
defines the {\it symmetry transformation}
${\mathcal L}_c\longrightarrow {\mathcal L}_c:[v]\mapsto [gv]$ 
of the carbon nanotube ${\mathcal L}_c$ if 
\[ [v]=[u]\Longrightarrow [gv]=[gu]\]
that is, if
\[ v-u\in \mathbb{Z}c\Longrightarrow gv-gu\in \mathbb{Z}c.\]
{\bf Theorem 2.} {\it The transformations
\begin{equation}\begin{array}{ll}
g_w:{\mathcal L}_c\longrightarrow {\mathcal L}_c\qquad  & g_w[v]=[v+w]\\
\tau :{\mathcal L}_c\longrightarrow {\mathcal L}_c & \tau [v]=[-v+\vartheta ]
\end{array}
\end{equation}
are symmetry transformation of ${\mathcal L}_c$ for all} $w\in {\mathcal T}$.\\[2mm]
{\bf Proof.} We have 
$v-u\in \mathbb{Z}c\Longrightarrow (v+w)-(u+w)=v-u\in \mathbb{Z}c$, and 
$v-u\in \mathbb{Z}c\Longrightarrow (-v+\vartheta )-(-u+\vartheta )
=u-v\in \mathbb{Z}c.$ \qquad  \rule{2mm}{2mm}

Let $n={\rm gcd}\{c_0,c_1,c_2\}$ be the greatest common divisor
of $c_0,c_1,c_2,$ and let $c=nc'$, that is, 
$c'_0=c_0/n$, $c'_1=c_1/n$ and $c'_2=c_2/n$.
The transformation $g_{c'}$ represents a rotation of
angle $2\pi /n$ of the nanotube with respect to its axis.
Since $(c_1-c_2)c_0+(c_2-c_0)c_1+(c_0-c_1)c_2=0$, 
the vector $w=(c_1-c_2,c_2-c_0,c_0-c_1)$ is orthogonal to $c$, and the 
corresponding transformation $g_w$ is a pure translation, that is, a
translation in the direction of the nanotube symmetry axis. The vector 
$b=(1/\mathcal{R} )(c_1-c_2,c_2-c_0,c_0-c_1)$, where 
\begin{equation}
\mathcal{R} ={\rm gcd}\{ c_1-c_2,c_2-c_0,c_0-c_1\} 
=\left\{ \begin{array}{lll}
n & {\rm if} & c'_1-c'_2\not\in 3\mathbb{Z}\\
3n & {\rm if} & c'_1-c'_2\in 3\mathbb{Z}
\end{array} \right. 
\end{equation}
defines the shortest pure translation of ${\mathcal L}_c$. 

From $c_0+c_1+c_2=0$ we get 
$(c_1-c_2)^2+(c_2-c_0)^2+(c_0-c_1)^2=3(c_0^2+c_1^2+c_2^2)$, that is, 
$\mathcal{R}^2||b||^2=3||c||^2$, whence
\begin{equation}
q=\frac{1}{\mathcal {R}}(c_0^2+c_1^2+c_2^2)\in n\mathbb{Z}.
\end{equation}
For any $w\in \mathcal {T}$ the projections of $w$ on $c$ and $b$
can be written as 
\[
\frac{\langle w,c\rangle }{||c||^2}c
=\left( w_1\frac{c_1-c_0}{\mathcal{R}}+w_2\frac{c_2-c_0}{\mathcal{R}}\right)
\frac{c}{q}\qquad 
\frac{\langle w,b\rangle }{||b||^2}b
=\left( w_1\frac{c_2}{n}-w_2\frac{c_1}{n}\right)\frac{b}{q'}
\]
where $q'=q/n$. It is well-known that in the case of two integer numbers 
$\eta , \mu \in \mathbb{Z}$ with ${\rm gcd}\{\eta ,\mu \}=1$ there exist 
$\alpha , \beta \in \mathbb{Z}$ with $\alpha \eta +\beta \mu =1.$
Since ${\rm gcd}\{(c_1-c_0)/\mathcal{R},(c_2-c_0)/\mathcal{R}\}=1$ and 
${\rm gcd}\{ c_2/n,c_1/n\}=1$ it follows that the projection of $\mathcal{T}$
on $c$ is $\mathbb{Z}c/q$ and the projection of $\mathcal{T}$ on
$b$ is $\mathbb{Z}b/q'$. Let $\omega \in \mathcal{T}$ be the shortest
vector with 
\begin{equation} 
\frac{\langle \omega ,b\rangle }{||b||^2}b=\frac{b}{q'}.
\end{equation}

If $\mathcal{L}_c$ is a chiral nanotube then its symmetry group $G_c$ is 
generated by the transformations $g_{c'}$, $g_\omega $ and $\tau $
(additional symmetry operations, namely, mirror and glide planes occur only
in the case of armchair and zig-zag nanotubes).
More than that, for any $[v]\in \mathcal{L}_c$ there exist $s\in \mathbb{Z}$,
$m\in \{ 0,1,...,n-1\}$ and $p\in \{ 0,1\}$ uniquely determined such
that
\begin{equation}
[v]=\tau ^pg_\omega ^sg_{c'}^m[0,0,0].
\end{equation}
The usual description of the atomic positions of the atoms forming a carbon
nanotube \cite{Vu,W} is based on this remark, and the set 
\begin{equation}
\{ (s,m,p)\ |\ s\in \mathbb{Z},\ m\in \{ 0,1,...,n-1\},\ p\in \{ 0,1\}\}
\end{equation}
is used as a mathematical model.

The subgroup $\tilde{G}_c$ of $G_c$ generated by $g_{c'}$ and $g_\omega $ is a 
commutative index-two subgroup, and $g_{c'}^n=I$, $g_\omega ^{q'}=g_b$, where 
$I$ is the transformation $I[v]=[v]$ for all $[v]\in \mathcal{L}_c$.
 The irreducible representations of $\tilde{G}_c$ 
are one-dimensional and can be described in terms of generators as \cite{Da}
\begin{equation}
\left\{ \begin{array}{l}
T_{(\kappa ,m)}(g_{c'})={\rm e}^{{-\rm i}2\pi m/n}\\
T_{(\kappa ,m)}(g_\omega )={\rm e}^{{-\rm i}\kappa a/q'}
\end{array} \right. \qquad {\rm where} \quad 
\left\{ \begin{array}{l}
m\in \{ 0,1,...,n-1\}\\
\kappa \in [0,2\pi q'/a)
\end{array} \right. 
\end{equation}
The irreducible representations of $G_c$ can be obtained from the irreducible 
representations of $\tilde{G}_c$ by using the index-two subgroup induction.

Consider the Hilbert space
$(l^2({\mathcal L}_c),\langle , \rangle )$, where
\begin{eqnarray} 
l^2({\mathcal L}_c)&=&\left\{ \psi :{\mathcal L}_c\longrightarrow \mathbb{C}\left| 
\sum_{v\in {\mathcal L}_c}|\psi (v)|^2<\infty \right. \right\} \nonumber \\
\langle \psi _1,\psi _2\rangle &=&
\sum_{v\in {\mathcal L}_c}\overline{\psi }_1(v)\psi _2(v) 
\end{eqnarray}
and the unitary representation of $G_c$ in $l^2({\mathcal L}_c)$ defined by 
\begin{equation}
g:l^2({\mathcal L}_c)\longrightarrow l^2({\mathcal L}_c)\qquad 
(g\psi )[v]=\psi (g^{-1}[v]).
\end{equation}

If $\varepsilon $ is a real number and  $\gamma _0, \gamma _1,\gamma _2$ are
three complex numbers, then the linear operator
\begin{equation}\label{Hc}
H:l^2({\mathcal L}_c)\longrightarrow l^2({\mathcal L}_c)\qquad
(H\psi )[v]=\varepsilon \psi [v]+\sum_{j=0}^2\gamma (v,v^j)\, \psi [v^j]
\end{equation}
with $\gamma (v,v^j)$ defined by (\ref{gamma}), 
is a self-adjoint operator.\\[2mm]
{\bf Theorem 3.} {\it For any $k=(k_0,k_1,k_2)\in {\mathcal E}$ satisfying 
the relation
\begin{equation}\label{kperpc}
\langle k,c\rangle =k_0c_0+k_1c_1+k_2c_2\in (2\pi /a)\mathbb{Z}
\end{equation}
the real numbers
\begin{eqnarray}
E_{\pm }(k)=\varepsilon \pm
|\gamma _0 {\rm e}^{ {\rm i}k_0a}+\gamma _1  {\rm e}^{ {\rm i}k_1a}+
\gamma _2  {\rm e}^{ {\rm i}k_2a}| \label{Ekc}
\end{eqnarray}
belong to the spectrum of} $H.$\\[2mm]
{\bf Proof.} For $k\in \mathcal{E}$ satisfying (\ref{kperpc}) we have
\[ v-u\in \mathbb{Z}c\quad \Longrightarrow \quad \varphi (v)\, {\rm e}^{{\rm i}\langle k,v\rangle a}
=\varphi (u){\rm e}^{{\rm i}\langle k,u\rangle a} \]
and hence the Bloch type function
\begin{equation} 
\psi _k :{\mathcal L}_c\longrightarrow \mathbb{C}\qquad 
\psi _k[v]=\varphi (v)\, {\rm e}^{{\rm i}\langle k,v\rangle a}
\end{equation}
is well-defined. If $E\in \{E_+(k),E_-(k)\}$ then there exists a non-null function
of this form satisfying the relation $H\psi _k =E\psi _k.$\qquad \rule{2mm}{2mm}

If $\gamma _0$, $\gamma _1$, $\gamma _2$ are real then $H$ is $G_c$-invariant.
Indeed, since $g_{c'}v^j=(g_{c'}v)^j$, $g_\omega v^j=(g_\omega v)^j$ and 
$\tau v^j=(\tau v)^j$ we have $gHg^{-1}=H$, for any $g\in G_c$.
Denoting  $\langle k,c\rangle =2m\pi /a$ and $\langle k,\omega \rangle =\kappa /q' $ we get
\[ (g_{c'}\psi _k)[v]=\varphi (v)\, {\rm e}^{{\rm i}\langle k,v-c'\rangle a}
={\rm e}^{-{\rm i}\langle k,c'\rangle a}\psi _k[v]
={\rm e}^{-{\rm i}2m\pi /n}\psi _k[v]\]
\begin{equation}
 (g_\omega \psi _k)[v]=\varphi (v)\, {\rm e}^{{\rm i}\langle k,v-\omega \rangle a}
={\rm e}^{-{\rm i}\langle k,\omega \rangle a}\psi _k[v]
={\rm e}^{-{\rm i}\kappa a/q'}\psi _k[v]
\end{equation}
\[ (\tau \psi _k)[v]=\varphi (\tau v){\rm e}^{{\rm i}\langle k, -v+\vartheta \rangle a}=
(\varphi (\tau v)/\varphi (v)) \,{\rm e}^{{\rm i}k_0a}\psi _{-k}[v]\]
that is, the subspace generated by $\psi _k$ and $\psi _{-k}$ is $G_c$-invariant.
More than that, these relations allow us \cite{Da} to classify the eigenstates $\psi _k$ by
using the quantum numbers $m$ and $\kappa $. The set (figure 2) 
\begin{equation}
\mathcal{B}_c=\left\{ k\in \mathcal{E} \ \left| \ \begin{array}{l}
\langle k,c \rangle =2m \pi /a\ \ {\rm with} \ \ m\in \{ 0,1,...,n-1\}\\
\langle k,\omega \rangle =\kappa /q' \ \ {\rm with} \ \ \kappa \in [0,2q'\pi /a)
\end{array} \right. \right\} .
\end{equation}
contains a $k$ corresponding to each class. The conduction $m$-band 
\begin{equation}
\{ E_+(k)\ |\ \langle k,c\rangle =2m\pi /a,  
\ \langle k,\omega \rangle \in [0,2\pi q'/a)\}
\end{equation}
 and the valence $m$-band
\begin{equation}
\{ E_-(k)\ |\ \langle k,c\rangle =2m\pi /a,  
\ \langle k,\omega \rangle \in [0,2\pi q'/a)\}
\end{equation}
can easily be determined by using (\ref{Ekc}). A graphic representation of these
conduction bands can be found in \cite{Da}.

The relation (\ref{kperpc}) defines a family of equidistant straight lines orthogonal 
to $c$ with the distance between neighbouring lines equal to $\delta =2\pi /(a||c||).$
Since the length of the projection of the vector
$(2\pi /(3a),-2\pi /(3a),0)$ on $c$ is 
$2\pi (c_0-c_1)/(3a||c||)=(c_0-c_1)\delta /3$,
the $K$ points belong to the straight lines (\ref{kperpc}) 
if and only if $c_0-c_1\in 3\mathbb{Z}$.

The Hamiltonian used in the tight-binding description of $\pi $ bands
in ${\mathcal L}_c$, with only first-neighbour $C-C$ interaction, has the
form (\ref{Hc}). Except for very small diameter nanotubes, the constants
$\gamma _0$, $\gamma _1$, $\gamma _2$ keep almost the same values as in the
case of a graphene sheet \cite{A,W}. Neglecting the effects 
of the curvature of the graphite sheet, we can assume
$\varepsilon =0$, $\gamma _0=\gamma _1=\gamma _2=\gamma \in (0,\infty )$, and 
we get the energy levels
\begin{equation} \label{Epm}  
E_\pm (k)\!=\!\pm \gamma  \sqrt{3+2\cos(k_0\!-\!k_1)a+2\cos(k_1\!-\!k_2)a+2\cos(k_2\!-\!k_0)a}.
\end{equation} 
From the form of the surface $E(k)$ \cite{S2}, it follows that the gap between
the valence an conduction bands is given by 
\begin{equation} 
\Delta E_c=\min_{k\in {\mathcal B}_c}E_+(k)-\max_{k\in {\mathcal B}_c}E_-(k)=
2\min_{k\in {\mathcal B}_c}E(k)
\end{equation} 
and we have $\Delta E_c=0$ if and only if $c_0-c_1\in 3\mathbb{Z}$. Therefore \cite{H,W},
the nanotube ${\mathcal L}_c$ is a conductor if $c_0-c_1\in 3\mathbb{Z}$, and a
semiconductor if $c_0-c_1\not\in 3\mathbb{Z}$. The minimum 
$\min_{k\in {\mathcal B}_c}E(k)$ is achieved \cite{S2} at a point 
lying on a straight line (\ref{kperpc}) in the vicinity of a point $K$. 

In the case of a magnetic field parallel to the nanotube axis the Hamiltonian
also has the form (\ref{Hc}), but 
$\gamma _0=\gamma  {\rm e}^{ {\rm i}\beta c_0a}$, 
$\gamma _1=\gamma  {\rm e}^{ {\rm i}\beta c_1a}$, 
$\gamma _2=\gamma  {\rm e}^{ {\rm i}\beta c_2a}$, 
where $\beta $ is a real constant describing the magnetic field strength
threading the nanotube \cite{L}. 
Choosing $\varepsilon =0$, $\gamma \in (0,\infty )$,
we get the energy levels
\begin{eqnarray}
E_{\pm }^{mag}(k)=\pm \gamma 
|e^{ {\rm i}(k_0+\beta c_0)a}+ {\rm e}^{ {\rm i}(k_1+\beta c_1)a}+ {\rm e}^{ {\rm i}(k_2+\beta c_2)a}| 
=E_\pm(k+\beta c)
\end{eqnarray}
for any $k$ satisfying (\ref{kperpc}). From this relation it follows that the presence
of a magnetic field parallel to the nanotube axis can modify drastically
the electronic properties of the nanotube, and this modification depends
essentially on the nanotube chirality \cite{L}.

\section{Concluding remarks}

The alternate description presented in this paper offers certain formal
advantages and is significantly different from the usual one. We think 
that it can be used as a complementary description, and may stimulate the 
interest of mathematicians in the fascinating geometry of carbon nanotubes.
We have re-obtained some known results \cite{Da,D,J,L,R,S2,W}
concerning the chiral carbon nanotubes 
in order to illustrate the proposed approach. 

Since our model is a factor space, we have to verify the independence of the 
representative $(v_0,v_1,v_2)$ we choose for $[v_0,v_1,v_2]$ in the case of 
any mathematical object we consider on $\mathcal{L}_c$. This is not an 
inconvenience for our approach but a useful criterion when we look for 
mathematical objects with possible geometric or physical significance .

\end{document}